\documentstyle[fleqn,epsfig]{revp}
\newfont{\sff}{cmssi12} 
\newfont{\bigsf}{cmss12 scaled 2000} 
\newfont{\midsf}{cmss12 scaled 1000} 
\newfont{\smlsf}{cmss12 scaled 600}  
\newfont{\bigsff}{cmssi12 scaled 2000} 
\newfont{\sfi}{cmssi10} 

\newcommand{\pmonth}{August, 1999}
\newcommand{\spec}{Nuclear Physics}
\newcommand{\no}{10}

\begin{document}
\parindent 0pt
\parskip 12pt
\setcounter{page}{1}

\title{Skyrme mean-field studies of nuclei 
far from the stability line}

\author{P.-H. Heenen,$^{*1}$ P. Bonche,$^{*2}$, S. \'Cwiok$^{*3}$,
 W. Nazarewicz$^{*4-*6}$ and A. Valor$^{*1}$ \\ \\  
\sff
*1 Physique Nucl\'{e}aire Th\'{e}orique, University of Brussels, Belgium\\
\sff
*2 SPhT, CEA, Saclay, France. \\
\sff
*3 Institute of Physics, Warsaw University of Technology, Poland \\
\sff
*4 Department of Physics, University of Tennessee, Knoxville, USA \\
\sff
*5 Physics Division, Oak Ridge National Laboratory, USA \\
\sff
*6 Institute of Theoretical Physics, University of Warsaw, Poland
             }

\abst{
Two applications of mean-field calculations based
on 3D coordinate-space techniques
are presented. The first concerns the 
structure of odd-$N$ superheavy elements
that have been recently observed experimentally
and shows the ability of the method to describe, 
in a self-consistent way, very heavy odd-mass nuclei. 
Our results are consistent with the experimental data.
The second application concerns the introduction
of correlations beyond a mean-field approach 
by means of projection techniques and  configuration mixing.
Results for Mg isotopes demonstrate that
the restoration of rotational symmetry plays
a crucial role in the description of $^{32}$Mg.
}

\maketitle
\thispagestyle{headings}

\section*{Introduction}

Mean-field calculations on a 3-dimensional mesh were
introduced in nuclear physics in the late 1970s
in the  studies of nuclear collisions within
the time-dependent Hartree-Fock method$^{1)}$.
In these early applications, schematic nuclear interactions
were used, without
spin-orbit and Coulomb terms, and applications
were limited to collisions between light nuclei.
The same technique 
has later  been applied
in a systematic study of the spectrocopic properties
of Zr and neighboring isotopes$^{2)}$.
New developments have then permitted the use 
of effective
interactions of the Skyrme type, including Coulomb and spin-orbit terms,
and  pairing correlations were considered
 the BCS approximation
with a seniority force. There are several advantages of  the
coordinate-space
technique which make it very attractive.
It allows one to describe very general intrinsic deformation
 with a good numerical accuracy.
It also provides
a very flexible computation  scheme, suitable for generalizations
(e.g., introduction of correlations beyond the mean-field theory).
The main developments of the method performed 
after the completion of the  Zr study
were the introduction of a cranking constraint
to describe rotating nuclei$^{3)}$, the introduction
of the configuration mixing using the generator coordinate method$^{4)}$,
the particle-number projected HF+BCS calculations$^{5)}$,
 and improvements in the
treatment of the pairing correlations by the use
of the HFB method$^{6)}$ with density-dependent
zero-range interactions$^{7)}$. Let us also mention
a theoretical study of the properties of coordinate-space
calculations$^{8)}$ which introduced
 powerful ways to interpolate
wave functions discretized on a 3D mesh.

 In this note, we shall first present a very representative
application of the method to  odd-N superheavy
elements$^{9)}$ (SHE). We shall then describe new 
developments$^{10)}$ 
related to the restoration of
 the rotational and particle number symmetries; these improvements
are expected to be 
extremely useful in  future studies 
of medium-mass nuclei
far from stability.

\section*{Structure of Odd-$N$ Superheavy Elements}

There is no consensus among theorists with regard to 
the  center of the shell stability in the 
region of spherical SHE.
For the neutrons, most calculations predict
a magic gap  at  $N$=184. 
However, because of different treatments of
the large Coulomb potential and spin-orbit interaction,
various models yield different predictions for
the position of the  
magic proton gap (cf. discussion in Ref.~$^{11)}$).
In this context, the synthesis
of the $N$=175 isotones reported by the 
 Dubna/Livermore (DL) collaboration
($Z$=114) $^{12)}$  and 
the Berkeley/Oregon (BO) team ($Z$=118) $^{13)}$
is particularly important.
Due to the large neutron number of the compound nucleus,
 the observed $\alpha$ chains do not
terminate at some known systems. Consequently, 
until the charge-mass identification is made,
one can use   theoretical arguments
to support/disprove the experimental assignment. 

We have performed a large-scale self-consistent
study  of properties of even-even, $A$-odd,
and odd-odd heaviest  nuclei. 
In our work, we have used
the Hartree-Fock-Bogoliubov (HFB) method with a 
Skyrme interaction in the particle-hole (ph)
channel and a delta force in the pairing channel. The details of 
the calculations closely
follow  Ref.~$^{11)}$.
The HFB   equations have 
been solved in the coordinate space according to the method of
 Ref.~$^{7)}$.
In the ph  channel,
the Skyrme effective interaction SLy4  has been used.
This parameterization$^{14)}$ has, 
in particular, been adjusted to
reproduce  long
isotopic sequences; hence, one  can expect it to
have good isospin properties.
The pairing strengths of the interaction  have then been
 adjusted to reproduce
the average proton and neutron pairing gaps
in even-even nuclei around $^{254}$Fm. The properties of
 one-quasiparticle states were calculated by means of the self-consistent
blocking.

\begin{table}
\caption{Predicted structure of 
the lowest one-quasi\-particle excitations
in the $^{289}$114 $\alpha$-decay chain. Each excitation
is characterized by the Nilsson quantum numbers of the
odd neutron, excitation energy, and quadrupole deformation.
The number in parentheses indicates the $Q_\alpha$ value for
the ground-state to ground-state transition.}
\begin{tabular}{clrc}
 Nucleus    & Orbital & Energy (MeV) & $\beta_2$ \\
  \hline
$^{289}$114 & [707]$\frac{15}{2}^-$ & 0     & 0.12 \\
  (9.64)    & [611]$\frac{1}{2}^+$  & 0.52  & 0.11 \\
            & [604]$\frac{7}{2}^+$  & 0.79  & 0.13 \\
\hline
$^{285}$112 & [611]$\frac{1}{2}^+$  & 0     & 0.14 \\
  (8.88)    & [611]$\frac{3}{2}^+$  & 0.60  & 0.14 \\
            & [707]$\frac{15}{2}^-$ & 0.62  & 0.13 \\
            & [606]$\frac{11}{2}^+$ & 0.65  & 0.15 \\
            & [604]$\frac{9}{2}^+$  & 0.72  & 0.15 \\
\hline
$^{281}$110 & [604]$\frac{9}{2}^+$  & 0     & 0.19 \\
  (9.32)    & [606]$\frac{11}{2}^+$ & 0.07  & 0.19 \\
            & [611]$\frac{1}{2}^+$  & 0.12  & 0.18 \\
            & [611]$\frac{3}{2}^+$  & 0.59  & 0.17 \\
            & [613]$\frac{5}{2}^+$  & 0.65  & 0.17 \\
\hline 
$^{277}$108 & [611]$\frac{1}{2}^+$  & 0     & 0.21 \\
            & [604]$\frac{9}{2}^+$  & 0.04  & 0.20 \\
            & [613]$\frac{5}{2}^+$  & 0.31  & 0.21 \\
            & [716]$\frac{13}{2}^-$ & 0.36  & 0.21 \\
\end{tabular}
\end{table}

\begin{table}
\caption{Same as in Table~1, except for 
the $^{293}$118 $\alpha$-decay chain.}
\begin{tabular}{clrc}
 Nucleus    & Orbital & Energy (MeV) & $\beta_2$ \\
  \hline
$^{293}$118 & [707]$\frac{15}{2}^-$ & 0     & 0.11 \\
  (11.59)   & [611]$\frac{1}{2}^+$  & 0.16  & 0.10 \\
            & [602]$\frac{5}{2}^+$  & 0.84  & 0.12 \\
            & [604]$\frac{7}{2}^+$  & 0.98  & 0.10 \\
\hline
$^{289}$116 & [611]$\frac{1}{2}^+$  & 0     & 0.13 \\
  (10.18)   & [606]$\frac{11}{2}^+$ & 0.62  & 0.14 \\
            & [611]$\frac{3}{2}^+$  & 0.66  & 0.13 \\
            & [604]$\frac{9}{2}^+$  & 0.69  & 0.14 \\
            & [707]$\frac{15}{2}^-$ & 0.72  & 0.13 \\
\hline
$^{285}$114 & [606]$\frac{11}{2}^+$ & 0     & 0.16 \\
  (10.60)   & [611]$\frac{1}{2}^+$  & 0.04  & 0.16 \\
            & [611]$\frac{3}{2}^+$  & 0.15  & 0.16 \\
            & [604]$\frac{9}{2}^+$  & 0.16  & 0.16 \\
\hline 
$^{281}$112 & [611]$\frac{1}{2}^+$  & 0     & 0.19 \\
  (10.85)   & [604]$\frac{9}{2}^+$  & 0.07  & 0.19 \\
            & [611]$\frac{3}{2}^+$  & 0.37  & 0.19 \\
            & [613]$\frac{5}{2}^+$  & 0.41  & 0.19 \\  
\end{tabular}
\end{table}

The calculated one-quasiparticle structures
and the  g.s.-to-g.s.\ values of $Q_\alpha$
in the $^{289}114$  and $^{293}118$ $\alpha$-decay chains
are shown in Tables~1 and 2, respectively. 

According to our calculations, the g.s-to-g.s.
 $\alpha$ decays of  $^{289}114$ and  $^{293}118$
are structurally forbidden since the 
g.s.\ properties of parent and daughter nuclei differ dramatically.
The allowed transitions to the excited  [707]$\frac{15}{2}^-$ level in 
$^{285}112$ and $^{289}116$ correspond to $Q_\alpha$=9.0\ MeV
and $Q_\alpha$=10.9\ MeV, respectively, 
 and are  considerably lower than the
experimental values (DL: 9.9\ MeV; BO: 12.6\ MeV)
Consequently, the most probable
candidates for the first $\alpha$ transitions  are 
the 10.2\ MeV  ($^{289}114$) and 11.8\ MeV 
($^{293}118$) lines
 associated with the allowed 
[611]$\frac{1}{2}^+$$\rightarrow$[611]$\frac{1}{2}^+$ decays. 
By the same token,
the  [611]$\frac{1}{2}^+$  g.s.\ of $N$=173 isotones
is expected to decay to the excited [611]$\frac{1}{2}^+$ level in 
the $N$=171 daughters. For $^{285}112$,
the corresponding $Q_\alpha$ energy, 8.76 MeV, is very close to the
experimental energy of the second $\alpha$ transition,
8.84\ MeV,  reported by the
DL group. For the $^{289}116$
decay we obtain  $Q_\alpha$=10.14\ MeV; i.e., we underestimate the
energy of the second $\alpha$ particle (11.8\,MeV) from the BO experiment.
This is the worst deviation from experiment obtained in our
calculations.

The [611]$\frac{1}{2}^+$ level is also expected to be the ground state of
the $N$=169 isotones, and
our prediction for the allowed 
[611]$\frac{1}{2}^+$$\rightarrow$[611]$\frac{1}{2}^+$ decays  
is 9.4 MeV for $^{281}110$ (DL: 9.0\,MeV) and 
10.6\ MeV for $^{285}114$ (BO: 11.5\ MeV). 

For the $\alpha$ transitions in the 
$^{281}112$$\rightarrow$$\cdots$$\rightarrow$$^{265}104$ chain,
allowing the positive-parity Nilsson orbitals with very
similar quantum numbers,
we obtain the following values of  $Q_\alpha$:
10.7\ MeV (BO: 10.8\ MeV),
11.1\ MeV (BO: 10.3\ MeV),
9.9\ MeV (BO: 9.9\ MeV),
8.0-8.5\ MeV (BO: 8.9\ MeV). An alternative route 
is possible that involves $\alpha$ transitions between 
$j_{15/2}$ orbitals. Here, for the last two $Q_\alpha$
values we obtain 9.8\ MeV and 8.7\ MeV. In both cases
we obtain good agreement with BO data. 

The calculated
equilibrium deformations of one-quasiparticle states in $N$=175
isotones
are rather small ($\beta_2$$\approx$0.11), and they increase along the
$\alpha$-decay chain. This trend reflects
the influence of the  $N$=184 magic neutron gap for the heaviest systems
and the deformed $N$=162 gap for the lightest nuclei in the chain.
As far as $Q_\alpha$ values are concerned, 
our  calculations are consistent with
the recent experimental findings.

\section*{Correlations beyond mean field in Mg isotopes}

The cranking method is widely used in nuclear
spectroscopy to describe high-spin states. Applications
based on effective nuclear interactions have been
particularly successful in the description of
superdeformed rotational bands in several
regions of the mass table.
In the cranking method,
a rotational band is generated by the rotation of a deformed
intrinsic state. Since
cranking states are not eigenstates of the angular momentum,
this causes some problems in determining, e.g., transition 
rates in nuclei which are not very well deformed. 

Another limitation of the cranking model appears
in nuclei which are soft with respect to the variation of a collective
variable. In this case, one expects the
interference of the 
zero-point vibrational mode with
the rotational motion which leads to 
variations in the nuclear structure along the yrast line.

In ref~$^{5)}$, we have presented a method to
restore the particle number symmetry within the
HF+BCS theory that allows us to perform a
configuration mixing of  projected wave functions
in the direction of
selected collective
variables.
The method presented in this section
generalizes it by the inclusion of
a restoration of the rotational symmetry.
It  enables us to describe details
of collective spectra and
transition rates.

 The starting point of the method 
are wave functions $ |\Phi{_\alpha}\rangle$ 
generated by mean-field calculations 
with a constraint on a collective coordinate ${\alpha}$. 
 Wave functions with good angular momentum and particle
numbers,
\begin{equation}\label{first}
|\Phi,JM\alpha\rangle = \frac{1}{N_{x}}\sum_{K}g_K
 {\hat{P}^{J}_{MK}\hat{P}^{Z}\hat{P}^{N} |\Phi_{\alpha}\rangle},
\end{equation}
are obtained by means of the projection operators  $\hat{P}$.
In Eq.~(\ref{first}),
 $N_x$ is a normalization factor, depending 
on $x=J,M,N,Z,\alpha$.

Using the projected state $|\Phi,JM\alpha\rangle$
as a generating function,
 configuration mixing along the collective variable $\alpha$
is  performed for each angular momentum:
\begin{equation}
|\Psi,JM\rangle =\sum_{\alpha}f_{\alpha}^{JM}|
                              \Phi,JM\alpha\rangle.
\end{equation}
The weight functions $f_{\alpha}^{JM}$ 
are found by requiring
that the expectation value of the energy,
\begin{equation}\label{E10a}
E^{JM}={\langle{\Psi, JM}\vert\hat H\vert{\Psi, JM}\rangle
    \over\langle{\Psi, JM}\vert{\Psi, JM}\rangle},
\end{equation}
is stationary with respect to
an arbitrary  variation $\delta f_{\alpha}^{JM}$. 
This prescription leads to
the discretized Hill-Wheeler 
equation 
\begin{equation}\label{E11}
 \sum_{\alpha}({\mathcal H}^{JM}_{\alpha,\alpha'}
      -E^{JM}_k{\mathcal I}^{JM}_{\alpha,\alpha'})
          f^{JM,k}_{\alpha'}=0,
\end{equation}
in which the Hamiltonian kernel ${\mathcal H}^{JM}$ and the overlap kernel
${\mathcal I}^{JM}$ are defined as
\begin{equation}\label{E12}
{\mathcal H}^{JM}_{\alpha,\alpha'}=
     \langle{\Phi JM\alpha}\vert\hat H\vert{\Phi JM\alpha'}\rangle\quad,
{\mathcal I}^{JM}_{\alpha,\alpha'}=
     \langle{\Phi JM\alpha}\vert{\Phi JM\alpha'}\rangle.
\end{equation}
The kernels (\ref{E12}) are  obtained by integrating 
 the matrix elements between
rotated wave functions
over three
Euler angles and two gauge angles.
Besides these kernels, one can determine transition probabilities
between different eigenstates of the Hill-Wheeler equation.
This requires the calculation of matrix elements 
of electromagnetic operators.

Currently, in order to save computing time
and to test the method, certain symmetry restrictions have been
imposed on the mean-field wave functions. Namely,
they have been assumed to be  axially
symmetric and time-reversal invariant. In this way,
the integration over the Euler angles is limited to 
a single angle. Pairing correlations are treated in the
BCS approximation.

In the following examples,
a Skyrme force is used in the particle-hole channel and
a density-dependent zero-range interaction in
the pairing channel. As in ref$^{5)}$, 
in the calculation
of non-diagonal matrix elements,
the density dependence
of the Skyrme interaction is generalized
to a dependence on the mixed density.

{\it Application to $^{24}$Mg}

 The results shown in this section have been obtained using
the HF+BCS wave functions generated with an axial quadrupole constraint.
The Lipkin-Nogami prescription has been used to improve
the treatment of pairing correlations. 
 The variation of the energy as a function of prolate and 
oblate deformations
is plotted in Fig.~1 for the Sly4 Skyrme interaction 
and a surface pairing  interaction having strength 
$G$=1000\,MeV fm$^3$ for both neutrons and protons. 
In addition to the  prolate absolute minimum,
the mean-field curve presents a shoulder
at an  oblate deformation around 50\,fm$^2$.
The triple projection creates an oblate minimum at the
position of the shoulder for $J$ up to 6$^+$. For greater
values of $J$,
the weights of the intrinsic wave functions
 for deformations below
--200\,fm$^2$ are very small. Consequently, the projected energy curves 
do not exhibit any oblate mimima. 

For each value of the angular momentum, we have performed a
configuration-mixing calculation including quadrupole
moments between --350fm\,$^2$ and 450\,fm$^2$. 
This corresponds to intrinsic configurations excited by
about 30\,MeV with respect to the prolate minimum.
The spectrum that is
generated in this way (represented by bars) is plotted at 
 the quadrupole moment corresponding to the largest
component of the collective wave function. The value
of this quadrupole moment  
is very close to the minimum of the projected
energy curve.
Moreover, the energy of this minimum is only slightly modified
by the configuration mixing. The largest gain, 
$\sim$800\,keV, is obtained for the 0$^+$ state, but it is reduced at
higher spins.
Several excited states are found at low energy for each spin value.
Except for the second 0$^+$ and 10$^+$, the wave functions of yrare
states are peaked around the oblate secondary minimum.

In the same figure, theoretical 
transition probabilities along the yrast line
are compared with the data. In the 
 GCM calculation, 
the minima of projected energy curves have been used. 
The $B(E2;2_1^+\rightarrow 0_1^+)$ rate calculated
between the minima of the two curves is very close to the
experimental value. The configuration mixing causes
a spreading of the collective wave function on the quadrupole moment
and decreases the value of the $B(E2)$ rate. 
Since for spins greater than zero wave functions do not
have low quadrupole-moment components, the configuration mixing 
does not affect significantly the transition probabilities. Here,
the agreement between  calculations and  experiment
is excellent. One has to note, however, that the inclusion of
triaxiality is expected to modify these predictions to some extent.

{\it Application to $^{32}$Mg}

Altough $^{32}$Mg corresponds to the $N$=20 shell closure,
there is some  experimental evidence that it is 
deformed in its ground state. The excitation energy of its
first 2$^+$ state is only around 900\,keV, i.e., significantly lower
than in lighter Mg isotopes. Also the
$B(E2;2_1^+\rightarrow 0_1^+)$
value is unusually large. 

Figure 2  shows the calculated energy curves
 for $^{32}$Mg.
The ground state calculated in the HF+BCS approximation
is spherical (see discussion in ref.$^{15}$). 
However, unlike in the $^{24}$Mg case,
the restoration of broken symmetries modifies the topology
of the energy curve in a more dramatic way. Namely,
the resulting 0$^+$ ground state corresponds to
the projection of a deformed intrinsic state with a quadrupole
moment of 100\,fm$^2$. The minima for the higher-$J$ values
can be associated with even larger deformations. The configuration
mixing leads to small energy gains.
The collective 0$^+$ wave function is spread over a large range of
 quadrupole moments between --200\,fm$^2$ and 200\,fm$^2$,
while for other $J$-values the weights of the
HF+BCS wave functions have well-pronounced 
maxima around 200\,fm$^2$.

The calculated spectrum and  transition probabilities are 
compared to the experimental data in Fig.~2. 
It is seen that our calculations do not reproduce the
pronounced collectivity of  $^{32}$Mg. Compared to the data,
the
energy of the 2$^+$ is too high and the
$B(E2)$ value is too low. It should be noted, however, that
the delicate balance between the spherical minimum
and the  deformed intruder configuration in this nucleus is 
 sensitive to the effective interaction used$^{15)}$.

This research was supported in part by
the U.S. Department of Energy
under Contract Nos.\ DE-FG02-96ER40963 (University of Tennessee),
DE-FG05-87ER40361 (Joint Institute for Heavy Ion Research),
DE-AC05-96OR22464 with Lockheed Martin Energy Research Corp.\ (Oak
Ridge National Laboratory), the Polish Committee for
Scientific Research (KBN) under Contract No.~2~P03B~040~14,
the NATO grant CRG 970196, the Rector Grant of Warsaw University 
of Technology, and by the PAI-P3-043 of the Belgian
Office for Scientific Policy.

\section*{References}
\re
1) H. Flocard, S.E. Koonin and M.S. Weiss, 
Phys. Rev. {\bf C17},  1682 (1978)
\re
2)  P. Bonche, H. Flocard, P.-H. Heenen,S.J. Krieger and M.S. Weiss,
                Nucl. Phys. {\bf A443},  39 (1985)
\re
3) P. Bonche, H. Flocard and P.-H. Heenen,
    Nucl. Phys. {\bf A467}, 115 (1987) 
\re
4)P. Bonche, J. Dobaczewski, H. Flocard, P.-H. Heenen and J. Meyer,
Nucl. Phys. {\bf A510}, 466 (1990)
\re
5)P.-H. Heenen, P. Bonche, J. Dobaczewski and H. Flocard,
Nucl. Phys. {\bf A561}, 367 (1993)
\re
6)B. Gall, P. Bonche, J. Dobaczewski, H. Flocard and P.-H. Heenen,
Z. Phys. {\bf A348}, 183 (1994)
\re
7)J. Terasaki, P-H Heenen, P. Bonche, J. Dobaczewski and H. Flocard,
Nucl Phys {\bf A600}, 371 (1996)
\re
8) D. Baye and P.-H. Heenen, J. Phys. A  {\bf 19}, 2041 (1986)
\re
9) S. \'Cwiok, W. Nazarewicz and P.H. Heenen,
Phys. Rev. Lett. {\bf 83}, 1108 (1999)
\re
10) A. Valor, P.-H. Heenen and P. Bonche, in preparation.
\re
11) S. \'Cwiok {\em et al.},    Nucl. Phys. {\bf A611}, 211 (1996)
\re
12) Y.T. Oganessian {\em et al.}, submitted to Phys. Rev. Lett. 1999; JINR
  Preprint E7-99-53, Dubna 199
\re
13)V. Ninov {\em et al.}, Phys. Rev. Lett. {\bf 83}, 1104 (1999)
\re
14)E. Chabanat {\em et al.},
Nucl. Phys. {\bf A635}, 231 (1998).
\re
15)P.G. Reinhard  {\em et al.},
Phys. Rev. {\bf C60}, 014316 (1999).
\re

\begin{figure*}[H]
\centerline{\epsfig{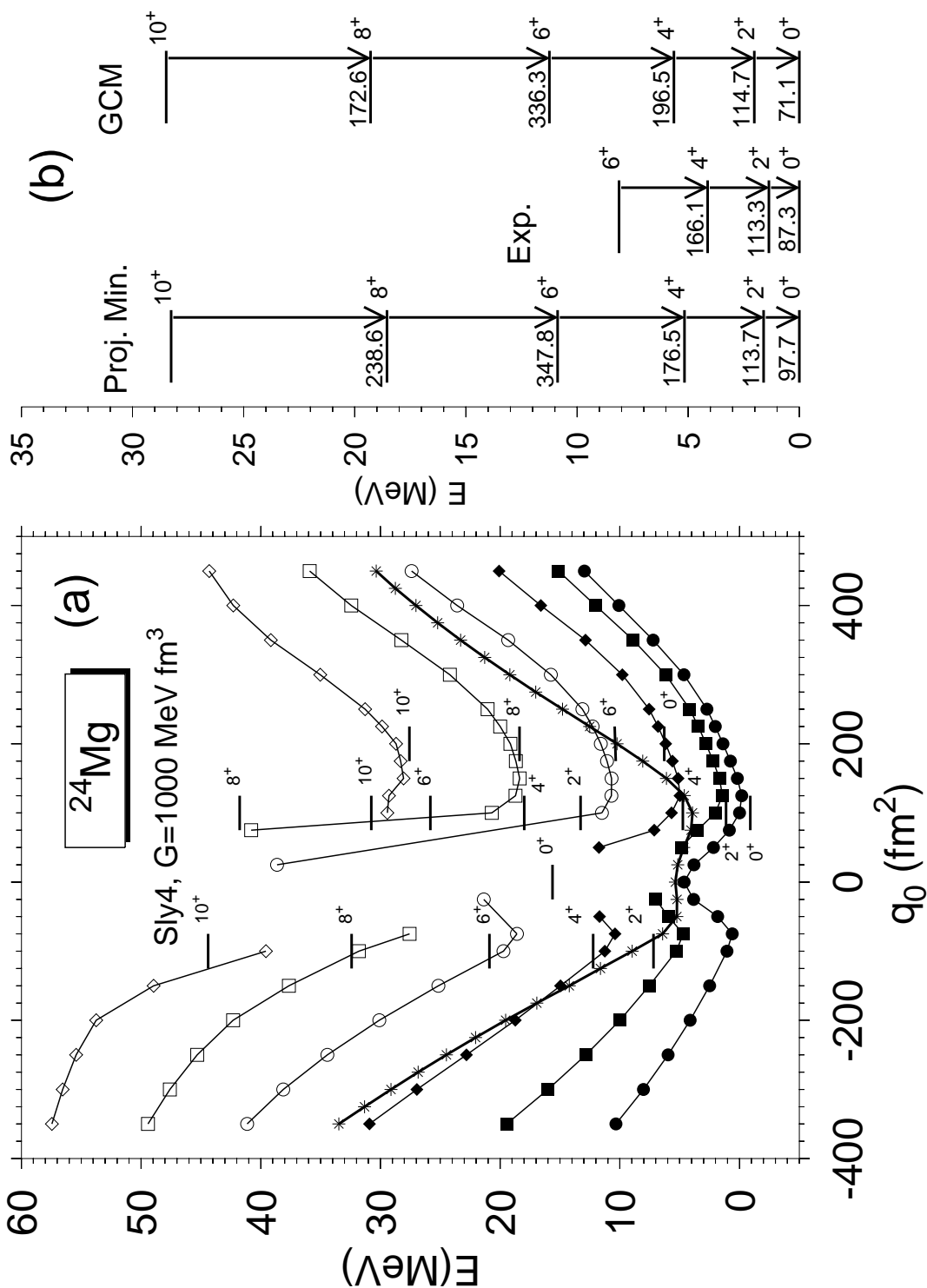}}
\caption{Symmetry-projected results for the nucleus $^{24}$Mg. 
(a) Energies as a function of the axial
quadrupole moment. Solid line with stars: 
HF+BCS+LN results; the
curves displaying filled circles, filled boxes,
filled diamonds, circles,  boxes, and
diamonds   correspond, respectively, to projected
energies in the $J$=0$^+$, 2$^+$, 4$^+$, 6$^+$, 8$^+$, 10$^+$ states.
The horizontal bars are the projected energies of the
collective intrinsic states. (b) Excitation
energies and transition
probabilities (numbers in e$^2$fm$^4$) between: intrinsic states
minimizing each projected energy curve (left), yrast collective states
(right), and experimental levels (center).}
\end{figure*}

\begin{figure*}[H]
\centerline{\epsfig{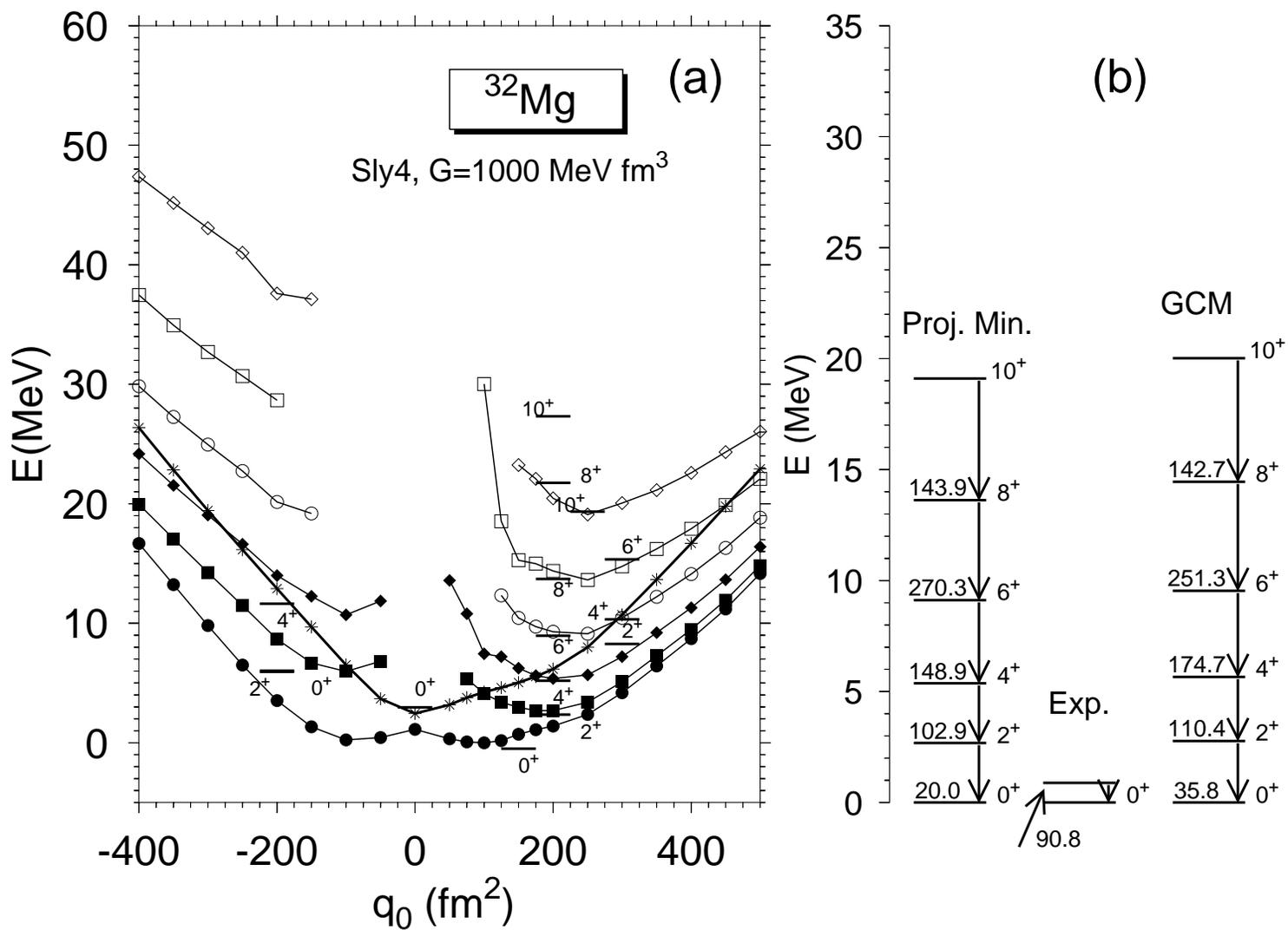}}
\caption{Same as in Fig.~1, except for
 $^{32}$Mg.} 
\end{figure*}

\end{document}